\documentclass{aa}
\usepackage{graphicx}

\begin{document}

\title{The gravitational torque of bars in optically
unbarred and
  barred galaxies\thanks{Based on observations made
with the
  William Herschel Telescope, operated on the island of La Palma
by
  the Isaac Newton Group in the Spanish Observatorio del Roque
de los
  Muchachos of the Instituto de Astrof\'\i sica de Canarias.}}

\author{D.L. Block\inst{1}
\and I. Puerari\inst{2}
\and J.H. Knapen\inst{3,4}
\and B.G. Elmegreen\inst{5}
\and R. Buta\inst{6}
\and S. Stedman\inst{4}
\and D.M. Elmegreen\inst{7}}

\offprints{D. L. Block}

\institute{Department of Computational and Applied Mathematics,
  University of the Witwatersrand, 1 Jan Smuts Avenue,
Johannesburg
  2001, South Africa\\
\email{igalaxy@iafrica.com}
\and Instituto Nacional de Astrof\'\i sica, Optica y Electr\'onica,
Calle Luis Enrique Erro 1, 72840 Tonantzintla, Puebla, M\'exico\\
\email{puerari@inaoep.mx}
\and Isaac Newton Group of Telescopes, Apartado 321, E-38700
Santa
  Cruz de La Palma, Spain \\
\email{knapen@ing.iac.es}
\and
University of Hertfordshire, Dept of Physical Sciences, Hatfield,
  Herts. AL10 9AB, United Kingdom\\
\email{knapen@star.herts.ac.uk and stedman@star.herts.ac.uk}
\and
IBM Research Division, TJ Watson Research Center, Box 218,
  Yorktown Heights NY 10598, USA\\
\email{bge@watson.ibm.com}
\and Dept of Physics and Astronomy, University of Alabama, Box
  870324, Tuscaloosa, Alabama 35487, USA\\
\email{buta@sarah.astr.ua.edu}
\and
Vassar College, Dept of Physics and Astronomy, Poughkeepsie,
NY
  12604-0278, USA\\
\email{elmegreen@vaxsar.vassar.edu`}}

\titlerunning{Gravitational Bar-torques}
\authorrunning{Block et. al.}

\date{Received .............../ Accepted 30 May 2001}

\abstract{The relative bar torques for 45 galaxies observed at $K$-
band
with
the 4.2m William Herschel Telescope are determined by
transforming the
light distributions into potentials and deriving the maximum ratios of
the tangential forces relative to the radial forces.  The results are
combined with the bar torques for 30 other galaxies determined
from our
previous $K$-band survey (Buta \& Block 2001).  Relative bar
torques
determine the degree of
spiral arm forcing, gas accretion, and bar evolution.  They differ from
other measures of bar strength, such as the relative amplitude of
the
bar
determined photometrically, because they include the bulge and
other
disk
light that contributes to the radial component of the total force. If
the
bulge is strong and the radial forcing large, then even a prominent
bar
can have a relatively weak influence on the azimuthal motions in
the disk.
Here we find that the relative bar torque correlates only weakly with
the
optical bar type listed in the Revised Shapley-Ames and de
Vaucouleurs
systems. In fact, some classically barred galaxies have weaker
relative
bar torques than classically unbarred galaxies.  The optical class is
a poor measure of azimuthal disk forcing for two reasons: some
infrared
bars are not seen optically, and some bars with strong bulges have
their
azimuthal forces so strongly diluted by the average radial force that
they exert only small torques on their disks.  The Hubble
classification
scheme poorly recognizes the gravitational influence of bars.
Applications of our bar torque method to the high-redshift universe
are briefly discussed.
}

\maketitle

\section{Introduction}

Bars have been recognized in galaxies since the time of Curtis
(1918) and Hubble (1926), but only recently have methods been
developed that quantify the impact of these features on galaxy
structure. Bar strength is important in galaxy
morphological studies because phenomena such as gas inflow,
angular
momentum
transfer, noncircular motions, lack of abundance gradients, nuclear
activity,
starbursts, and the shapes and
morphologies of rings and spirals, may all be tied in various ways
to
the effectiveness with which a bar potential influences the motions
of stars and gas in a galactic disk (e.g., Sellwood \& Wilkinson
1993; Buta \& Combes 1996; Knapen 1999).

In the past, bar strength was judged visually from galaxy images
on blue-sensitive photographic plates. Hubble (1926) divided
galaxies into barred (SBa, SBb, ...) and
normal (Sa, Sb, ...) spirals along a famous ``tuning fork.''
This view was revised by de Vaucouleurs (\cite{dev59}), whose
classification volume recognized apparent bar strength (SA, SAB,
SB)
as a continuous property of galaxies called the ``family''.
De Vaucouleurs' main contribution here was to recognize the
existence
of galaxies having a bar intermediate in apparent strength between
nonbarred and barred spirals. This is the essence of category SAB.

However, neither the Hubble nor the de Vaucouleurs bar
classifications
can be expected to be accurate measures of bar strength because
apparent bar strength is impacted by wavelength, the effects of
extinction
and star formation, inclination and bar orientation
relative to the line of sight, and also on observer interpretations.
It is well known that bars are more
prominent in near-infrared images than in blue-light images (e.g.,
Block \& Wainscoat 1991; Knapen,
Shlosman \& Peletier 2000; Eskridge
et al. 2000). Clearly, the near-infrared is the best wavelength
regime for judging bar strength.

In the near-infrared, one probes the older, star dominated disk.
This led Block and Puerari (1999) to propose a simple
classification scheme for spirals in the near-IR involving
the dominant Fourier $m$-mode and the pitch angle of the spiral
arms. Old disks may be grouped into three principal
archetypes: the tightly wound $\alpha$ class, an intermediate
$\beta$
class (with pitch angles of $\sim$ 25$^{\circ}$); and the $\gamma$
class, in which the pitch angles in the near-infrared are $\sim$
40$^{\circ}$.  Flat or falling rotation curves give rise to the tightly
wound
$\alpha$ class; rising rotation curves are associated with the open
$\gamma$ class. Hence, these dust penetrated classes are
inextricably
related to the rate of shear in the stellar disk (Block et al. 1999).

In this paper, we further describe the newest quantitative parameter
in our near-infrared classification
scheme: the gravitational torque of a bar embedded in its disk,
based on the theoretical definition of Combes \& Sanders
(\cite{combes81}). This parameter, called $Q_b$, has recently been
developed by Buta \& Block (\cite{buta01}). Our goal here is to
build on this study and to evaluate the visual bar classifications
of Hubble and de Vaucouleurs by comparing them with $Q_b$.

\section{Dust-Penetrated Classifications for 45 Galaxies}

We base our study on a combination of the sample used by Buta
\&
Block (2001) and analysis of a new sample of galaxies imaged
with the 4.2-m William Herschel
Telescope (WHT) using the Isaac Newton Group Red Imaging
Device,
INGRID.
The INGRID camera uses a 1024$\times$1024 HgCdTe
HAWAII array, optimised for imaging between 0.8 to 2.5$\mu m$.
The
scale is 0.242$''$ per pixel, giving a 4.13$'$ $\times$ 4.13$'$ field
of view. INGRID can be used to efficiently mosaic relatively large
galaxies in rather short amounts of observing time.
We used the $K_{\rm s}$ filter
(central wavelength 2.150$\mu m$) in securing the
observations
reported here.
Three galaxies (NGC~628, 6140 and 6946) were observed by R.S.
de Jong
using Steward Observatory's 2.3m Bok Telescope on Kitt Peak.

The galaxies were selected
to have an angular diameter greater than 4.2 arcminutes
and an inclination less than 50 degrees.
Our sample covers a wide range of Hubble types, de Vaucouleurs
class
(T index) as well as form family (SA, SAB
and SB). The Elmegreen arm class (Elmegreen \& Elmegreen
1987) for the sample spans the entire
range, from flocculent types 1-3 to `extreme grand design' class
12.
A discussion of the
sample appears in Stedman \& Knapen
(\cite{sted01}) and in Knapen et al. (in preparation). The latter paper
also gives a full description of the data gathering and reduction
procedures used.

Table 1 lists the galaxies observed.
This sample has the advantage in that it contains many
famous spirals whose angular diameters would normally be too
large to be included in other 4-m class near-infrared imaging
surveys.
Included are galaxies such as NGC 4501 (M88,
one of the largest spirals in the Virgo cluster) and Hubble Atlas
prototypes
such as NGC 628 (M74), NGC 3351 (M95) and NGC 4321 (M100).

A variety of quantitative parameters has been suggested or could
be interpreted to represent a measure of the strength of a bar,
as discussed by Buta \& Block (2001).
The simplest is the deprojected bar axis ratio, $(b/a)_{bar}$,
developed by Martin (1995) and listed in his Table as $(b/a)_i$.
(This can be expressed as a bar ellipticity index
$\epsilon_{bi} = 10[1-(b/a)_i]$.) Martin's parameter does not
depend on spectroscopic observations, surface photometry, or
mass-to-light ratio assumptions, but nevertheless should relate
to bar strength based on the analytical models of Athanassoula
(1992).
This kind of parameter
has also been used in a number of other recent papers
such as Rozas, Knapen \& Beckman (\cite{rozas98}),  Knapen,
Shlosman \&
Peletier (2000), and Abraham \& Merrifield (\cite{abr00}). A $K$-
band
Fourier
analysis of bar strength is discussed by Regan \& Elmegreen
(1997).
Martin (1995) notes that $(b/a)_i$ is not a complete description
of bar strength, but merely the most accessible one. The $Q_b$
parameter has the advantage in that it is not necessary to
rigorously
define the bar to measure its strength in this manner (i.e., where the
bar begins or ends), while bar axis ratios depend on what we see
as the bar.

\begin{figure}
\resizebox{\hsize}{!}{\includegraphics{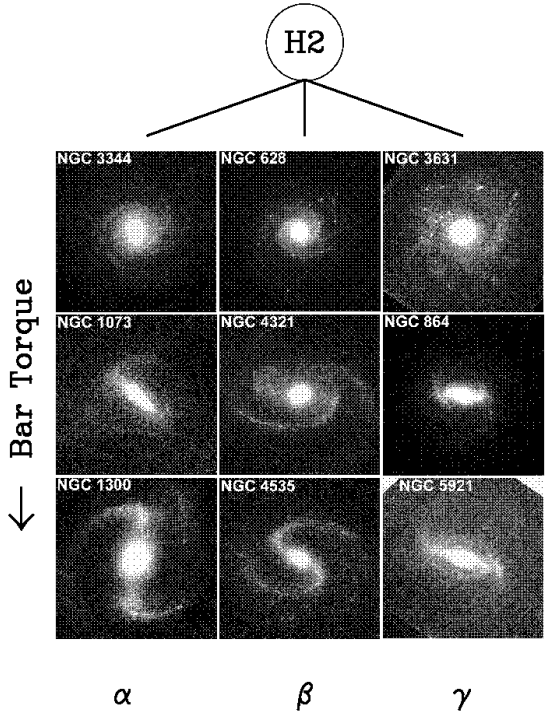}}
\caption{A selection of nine galaxies observed with the 4.2-m
William
  Herschel Telescope. All images have been deprojected. The
spirals
illustrated here are of class H2 i.e., they are two
  armed, and have a dominant m=2 Fourier
  harmonic. Their complete dust penetrated (DP) classifications
appear
  in column 4 of Table 1.}
\label{make_mosaic_class}
\end{figure}

\begin{figure}
\resizebox{\hsize}{!}{\includegraphics{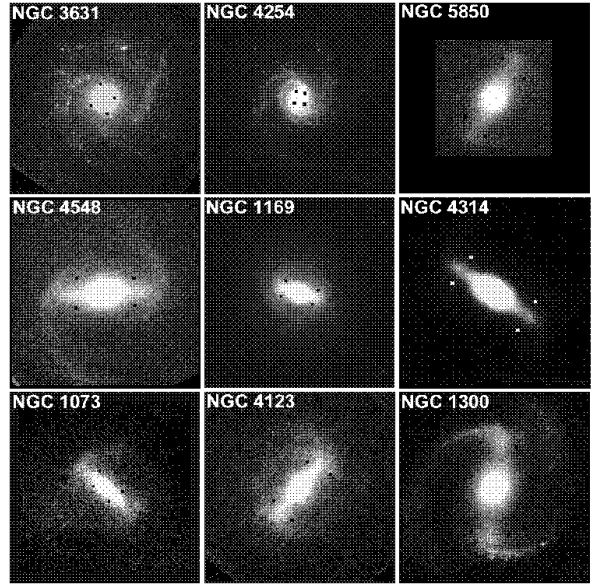}}
\caption{A sequence of galaxies (deprojected) ranked in terms of
increasing bar
  torque; individual torque classes are listed in column 4 of Table 1.
The four filled black or
yellow
  dots per image
indicate the
  locations of where the ratio of the tangential force to the mean
  axisymmetric radial force  reaches a maximum (in modulus), per
quadrant.}
\label{make_mosaic_dot}
\end{figure}

Dust-penetrated classifications, including principal harmonic H$m$,
arm pitch
angle class P$_c$, and bar torque class B$_c$, were derived for
each
galaxy in Table 1
following Block \& Puerari (\cite{block99}) and Buta \& Block
(2001), and are listed in column 4 in the form H$m$P$_c$B$_c$.
Fig. 1 shows our near-infrared tuning
prong for nine of the galaxies observed
using the WHT;  all galaxies illustrated in this Figure have
been deprojected to appear `face-on'.


The amplitude of each Fourier component is given by
(Schr\"oder et al., 1994):

$$\displaystyle A(m,p) = \frac{\Sigma_{i=1}^I \Sigma_{j=1}^J I_{ij}
({\rm ln}\;r,\theta)\; {\rm exp}\;(-i(m \theta + p\; {\rm
ln}\;r))}{\Sigma_{i=1}^I \Sigma_{j=1}^J I_{ij} ({\rm ln}\;r,\theta)}$$

\noindent where $r$ and $\theta$ are polar coordinates, $I({\rm
ln}\;r,\theta)$
is the intensity at position $({\rm ln}\;r, \theta)$, $m$ represents the
number of
arms or modes, and $p$ is the variable associated with the pitch
angle
$P$,
defined by $\displaystyle \tan P = -\frac{m}{p_{\rm max}}$.
In other words, to find the pitch angle of the dominant spiral
pattern or mode, we determine that value of $p$ which maximises
the
Fourier amplitude A(m,p) (see Block \& Puerari 1999 for a full
discussion).

It is important to note that in our formulation,
a logarithmic spiral (with a
single, well-defined pitch angle independent of radius) may be
generated for each mode $m$ (corresponding to the locale where A(m,p)
peaks). A set of possibly different pitch angles may be computed,
but we bin galaxies in Table 1 according to the `principal' pitch angle
corresponding to the dominant mode. However, we are fully aware that the
spiral
arms in barred galaxies often depart from a logarithmic shape. The arms
may break at a large angle to the bar and then wind back
to the other side, as in a ``pseudoring.'' Also, barred
spirals may have two spiral patterns, as in an ``inner
pseudoring'' and an ``outer pseudoring'' (e.g. NGC 3504;
see Sandage \cite{san61}).
These distortions show that the disk structure `feels' the
potential of the bar (Kormendy 1979, Kennicutt 1981).
We minimize the impact of rings and pseudorings by
excluding from our analysis the bar regions of the
galaxies in question. The
pitch angles listed in Table 1 correspond to arms
{\it outside} of the inner bar/bulge or inner pseudoring
region.

\begin{figure}
\resizebox{\hsize}{!}{\includegraphics{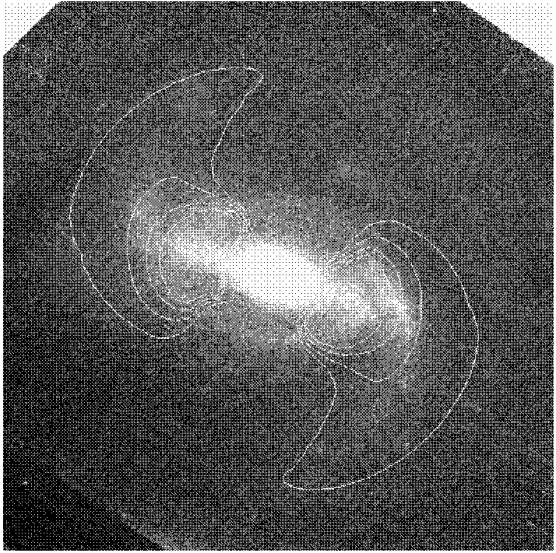}}
\caption{A deprojected near-infrared image of the ringed SB(r)
galaxy
NGC 5921 with inverse Fourier transform contours ($m$=2
harmonic)
superimposed.
The contours trace out two grand design stellar arms exterior to the
bar
and inner ring (see also Panel 206 in Sandage \& Bedke 1994).}
\label{make_mosaic_dot}
\end{figure}

To illustrate what the Fourier method is extracting
in a typical inner-ringed system in Table 1,
consider the SB(r) galaxy NGC 5921 (Figs. 1 and 3).
Fig. 3 shows the $m$=2 inverse
Fourier transform contours superposed on a deprojected image. The
contours trace out the two grand design arms extending outward
from the ring (see also panel 206 in `The Carnegie Atlas of Galaxies',
Sandage \& Bedke 1994). The Fourier method provides a reasonable
measure of the openness of the two outer arms in this case. In galaxies
with stronger rings, the dominant part of the outer arm pattern
should still provide a well-defined pitch angle.

In general, strong
outer rings and pseudorings have little impact on our classifications
because the
relative frequency of such features is fairly low. Only 10\% of RC3 disk
galaxies
within about 20 Mpc have such features (Buta \& Combes 1996), therefore
the classifications
for barred and nonbarred galaxies should have a comparable precision.
Our pitch angle classes in Table 1 are in any case a `first order'
approximation, set within the limits imposed by the distortions of
the arms in both normal and barred galaxies. As in the case of optical
classifications of barred galaxies, we find that from an operational
point of view, the dominant degree of openness of the arm pattern
for galaxies in Table 1 can be determined,
quantified and placed in one of our three broad classes $\alpha$,
$\beta$ or $\gamma$.




Bar torque classes are derived from
the maximum value of the ratio of the tangential force to the
mean axisymmetric radial force (Combes \& Sanders 1981)
using gravitational potentials
derived from the near-IR images under the assumptions of a
constant mass-to-light ratio and an exponential vertical
scale height (Quillen, Frogel, \& Gonz\'alez 1994).
Bar class B$_c$=1 includes galaxies having relative torques
$Q_b$= 0.1$\pm$0.05 (meaning the tangential force reaches a
maximum of 10\% of the axisymmetric background radial force);
class 2 involves those with $Q_b$= 0.2$\pm$0.05,
etc., up to class 6 (Buta \& Block 2001).
Uncertainties introduced in the torque
method are fully discussed in Buta \& Block (\cite{buta01}).
In cases where pitch angles for spiral arms could not be
determined due to low S/N for example, only the bar class number
is given for the dust penetrated (DP) class in Table 1.

The referee questions whether the bar strength code could be deceived
by two strong spiral arms in a nonbarred galaxy, giving a falsely
significant bar strength. In general, this is not likely.
Two-armed spirals betray their presence by peaks at either $p>0$
or $p<0$ in the $m$=2 mode, whereas bars reveal their presence by a peak
at
$p$=0 in the $m$=2 component.
{\it A relative maximum in A(2, 0) is the Fourier spectral indicator of
the presence of a bar}.
An independent check to ensure that one never confuses $m$=2
spirals with bars is
simply to look at the tangential-to-radial force ratio map:
in such a map, the characteristic
signature of a bar is the rectangular or parallelogram
pattern of the four maxima or minima (see Fig. 1 of Buta \& Block 2001).
It is always possible to overlay the locations of these points on
deprojected
images, as in Fig. 2, to make sure that they are not connected with the
spiral
arms.

The constancy of the mass-to-light ratio has not
been explicitly tested for any of the galaxies in Table 1 (e.g.,
by fitting an observed rotation curve or computing a near-IR color
distribution). However, based
on discussions by Freeman (1992) and Persic, Salucci, and Stel
(1996),
we believe the assumption
should be valid for the galaxies in our sample, whose average
luminosity is comparable to the Galaxy. (It would
not be valid for dwarf galaxies, none of which we study here).

Constraints on the dark halo content of barred galaxies
may also be deduced from the dynamical friction or drag of bars
rotating within dark matter halos. The studies of Debattista \&
Sellwood (\cite{deb98}, \cite{deb00}) indicate that bars are only
able to maintain
their high pattern speeds if the disk itself  provides
most of the gravitational potential; a high central density, dark
matter halo would simply provide too much drag on the bar.
Such independent studies suggest that light effectively traces
mass within the optical disks of barred spiral galaxies.

The vertical scaleheights of our preferentially face-on galaxies are
also not known. As noted by Buta \& Block (2001), this is one of
the
principal uncertainties in the $Q_b$ technique.
Our run of bar torque determinations in Table 1 assume  an
exponential
scaleheight for each galaxy of 325 parsecs
(in other words, equal to the exponential scaleheight of
our Galaxy; see Gilmore \& Reid \cite{gil83}).
Of course not all galaxies have the same exponential scaleheight:
the
study by de Grijs (\cite{deg98}) indicates that
late-type galaxies on average have a thinner disk than earlier type
systems.  To account for possible variations in scaleheight based
on morphological
type, and for the possibility that some bars are thicker than their
disks,
we have conducted separate potential runs for scaleheights $h_z$
= 225pc and
425pc.

The effect of varying the scaleheight from 225 to 425 pc is
to move a galaxy by one bar class, at most;
many galaxies retain their bar
classes with the three different scaleheight runs.
When a galaxy does move from one bar class to the next,
the effect of decreasing the scaleheight is to increase the bar
strength;
increasing the scaleheight leads to a decrease in the bar strength.
Buta \& Block (\cite{buta01}) find that an uncertainty of
$\pm$100pc in $h_z$
produces an average uncertainty of $\mp$13\% in bar strength.
In the future, it should be possible to improve our judgment of $h_z$
by scaling from values of the radial scalelength, as done by Quillen,
Frogel \& Gonz\'alez (\cite{qui94}).

Finally, in this preliminary analysis, we have assumed that each
bulge is as flat as the disk.  If the light distribution of a spherical
bulge is transformed into a potential assuming it is a thin disk, then
the axisymmetric radial forces derived will be too large, especially
in the bulge-dominated region where the error can reach a factor
of two. However, as discussed by Buta \& Block (2001),
as long as the bulge-dominated region is well inside the ends of the
bar, this effect will have little or no impact on the measured bar
strength.

\begin{table}
\caption{Optical and near-infrared classifications. Column 1
  lists the NGC number; column 2 the Hubble type extracted from
the
  Revised Shapley Ames Catalogue (Sandage and Tammann
\cite{san81});
column 3
the de Vaucouleurs (\cite{dev63}) form family;
column 4 lists our dust penetrated (DP)
  classification at K$_{s}$. The format used in column 4 is H$m$
(where
  $m$ is the dominant Fourier harmonic), followed by the pitch angle
  class ($\alpha$, $\beta$ or $\gamma$), followed by the bar torque
class.}
\begin{tabular}{cllccclcc}
\hline\noalign{\smallskip}

Galaxy & RSA type   & Form Family & DP Type

\\

\hline\noalign{\smallskip}

NGC0210  & Sb(rs)I          &   SAB  &   H2$\beta$1
\\
NGC0488  & Sab(rs)I         &    SA   & ...1
                                   \\
NGC0628  & Sc(s)I           &    SA   &   H2$\beta$0
\\
NGC0864  & Sbc(r)II-III     &   SAB    &  H2$\gamma$3

\\
NGC1042  & Sc(rs)I-II       &   SAB   &
H2$\beta$3
\\
NGC1073  & SBc(rs)II        &    SB &    H2$\alpha$4
\\
NGC1169  & SBa(r)I          &    SAB    &  H2$\gamma$3

\\
NGC1179  & SBc(rs)I-II      &   SAB    &   H2$\alpha$3
\\
NGC1300  & SBb(s)I.2        &  SB     &   H2$\alpha$5
\\
NGC2775  & Sa(r)            &   SA   & ...0
                                   \\
NGC2805  & ------           &   SAB   &
H2$\gamma$2
\\
NGC3184  & Sc(r)II.2        &   SAB  &
H2$\beta$1
\\
NGC3344  & SBbc(rs)I        &  SAB    &   H2$\alpha$1

\\
NGC3351  & SBb(r)II         &   SB     &  ...2
\\
NGC3368  & Sab(s)II         &  SAB     & ...2

\\
NGC3486  & Sc(r)I-II        &   SAB    &  H2$\gamma$0

\\
NGC3631  & Sc(s)I-II        &    SA     &   H2$\gamma$0

\\
NGC3726  & Sc(r)I-II        &    SAB    &  H2$\gamma$2

\\
NGC3810  & Sc(s)II          &   SA     &    H2$\beta$1
\\
NGC4030  & Sbc(r)I          &   SA     &   H2$\beta$1
\\
NGC4051  & Sbc(s)II         &   SAB   &
H2$\gamma$2
\\
NGC4123  & SBbc(rs)         &    SB     & ...4
\\
NGC4145  & SBc(r)II         &   SAB    &   H2$\beta$2
\\
NGC4254  & Sc(s)I.3         &   SA     &  H2$\gamma$1

\\
NGC4303  & Sc(s)I.2         &  SAB    &   H2$\gamma$3

\\
NGC4314  & SBa(rs) pec      &  SB     &   H2$\gamma$3

\\
NGC4321  & Sc(s)I           &  SAB    &  H2$\beta$2
\\
NGC4450  & Sab pec          &  SA     & ...2
                                  \\
NGC4501  & Sbc(s)II         &   SA   &   H2$\gamma$1

\\
NGC4535  & SBc(s)I.3        &  SAB   &
H2$\beta$3
\\
NGC4548  & SBb(rs)I-II      &   SB     &  H2$\gamma$2

\\
NGC4579  & Sab(s)II         &   SAB    &  H2$\gamma$2

\\
NGC4618  & SBbc(rs)II.2 pec &    SB     &  H1$\gamma$2

\\
NGC4689  & Sc(s)II.3        &    SA     &  H2$\beta$0
\\
NGC4725  & Sb/SBb(r)II      &   SAB    & ...3
           \\
NGC5247  & Sc(s)I-II        &   SA     &  H2$\gamma$1

\\
NGC5248  & Sbc(s)I-II       &  SAB   &
H2$\beta$0
\\
NGC5371  & Sb(rs)I/SBb(rs)I &  SAB   &
H1$\gamma$1
\\
NGC5850  & SBb(sr)I-II      &  SB     & ...2
                                 \\
NGC5921  & SBbc(s)I-II      &   SB     &  H2$\gamma$4

\\
NGC5964  & ------           &   SB     &    H2$\gamma$5
\\
NGC6140  & ------           & ---    &   H1$\gamma$--
\\
NGC6384  & Sb(r)I           &   SAB   &
H2$\beta$1
\\
NGC6946  & Sc(s)II          &   SA   &
H2$\gamma$0
\\
NGC7741  & SBc(s)II.2       &   SB   &  ...5
                                \\

\end{tabular}
\end{table}

The ratio of the tangential force to the mean axisymmetric radial
force reaches a maximum
or minimum around or near the ends of the bar.
Fig. 2 shows a
montage of nine WHT $K_{s}$ images wherein galaxies are ranked
in
terms of increasing gravitational bar torque.
The characteristic signature
of a bar in each of the images may be seen in Fig. 2 by noting
the location of the filled
black or yellow dots, where the ratio reaches a maximum or
minimum in each quadrant.

\section{Bar Torque and Form Family}

Combining the sample in Table 1 with that in Table 1 of Buta
and Block (2001), we have 75 galaxies for which $Q_b$ is now
available. Six of the galaxies in Table 1 are in common with
the list in Buta \& Block (2001). Except for NGC 4548, the values
are in good agreement, with differences attributable
to the quality of the images. We give preference in our analysis to
the
Table 1 values, because the INGRID images are superior in signal-
to-noise
to the images used by Buta \& Block (2001). In the case of NGC
4548,
a rebinning error caused Buta \& Block (2001) to overestimate the
bar
strength; the Table 1 value is the actual bar torque in this galaxy.

Fig. 4 shows $Q_b$ versus the Hubble S and SB classifications,
extracted
from the ``Revised Shapley-Ames Catalog'' (RSA; Sandage \&
Tammann
\cite{san81}). Fig. 5 is a similar
plot, but for the de Vaucouleurs (\cite{dev63})
classifications.\footnote{In some SAB classifications, de
Vaucouleurs
  (1963) underlined the A or the B to emphasize further subdivisions
  in this category. However, we ignore the underlines here since
there
  are too few galaxies in the underlined subdivisions in our sample.}
For a number of galaxies in Table 1, Martin (1995) lists an
estimate of the deprojected visual bar axis ratio, $(b/a)_{bar}$.
Fig. 6 shows a
plot of our gravitational bar torque $Q_b$ vs. $(b/a)_{bar}$.
Several points are noteworthy:

\begin{figure}
\resizebox{\hsize}{!}{\includegraphics{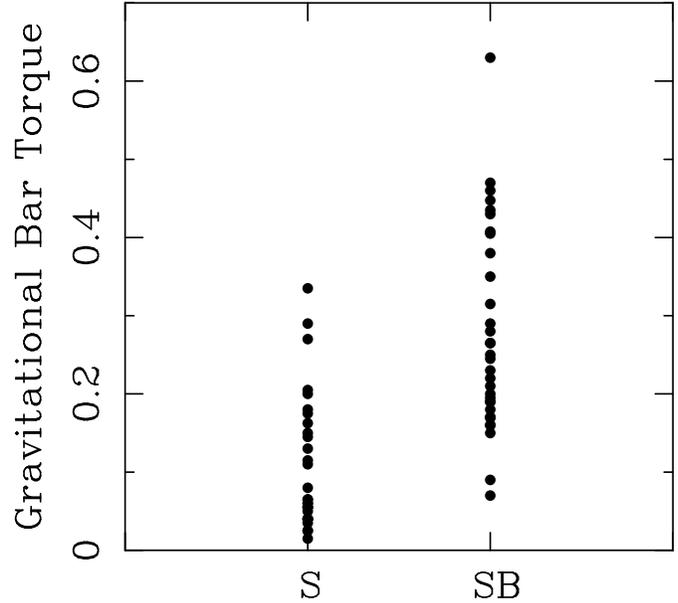}}
\caption{Bar torque versus Hubble Classification as prescribed in
the
  RSA for 64 galaxies, based on the combined sample from Table 1
  and the similar table in Block \& Buta (2001). Spirals classified as
Sa, Sab,
  Sb, Sbc etc are all grouped into the unbarred `S' bin; those of type
  SBa, SBab etc into the barred SB bin.}
\label{bartorqsasb}
\end{figure}

\begin{figure}
\resizebox{\hsize}{!}{\includegraphics{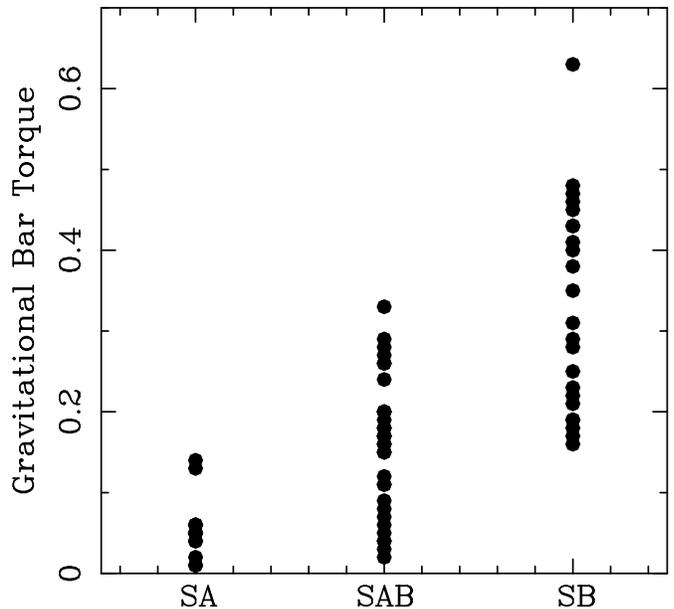}}
\caption{Bar torque versus the de Vaucouleurs (1963) form family for
69 galaxies, again based on the combined sample. The plot
excludes 6 galaxies whose form family is from other sources, as
well as NGC 4618, classified by de Vaucouleurs as a magellanic
barred spiral.}
\label{bartorqtype}
\end{figure}

\begin{figure}
\resizebox{\hsize}{!}{\includegraphics{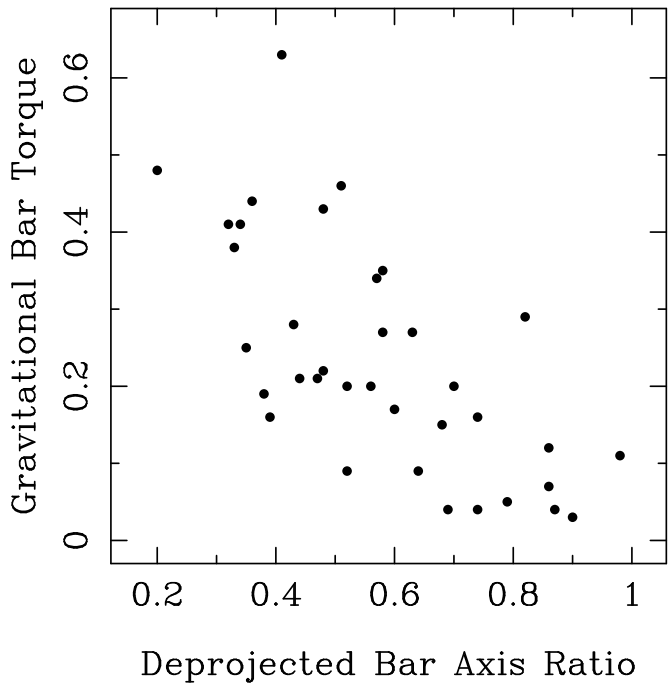}}
\caption{A comparison between gravitational bar torques $Q_b$
and
  deprojected bar axis ratio determined by Martin (1995). Highly
elongated bars
  have low values of $b/a_{bar}$, where a and b denote the bar major
  and minor axis respectively. An
  important point to note is that highly elongated `strong' bars
in the definition of
  Martin (1995) may have weak gravitational bar torques}.
\label{bartorqmartin}
\end{figure}

(i) Category S in Fig. 4 includes
galaxies ranging from bar torque class 0 (e.g., NGC 628)
to bar class 3 (e. g., NGC 1042).  NGC 4321, a Hubble Sc
prototype,
has a bar class of 2. Likewise, NGC 4450 (Sab) is of bar class
2.
NGC 4450 is illustrated in Panel 110 of Sandage \& Bedke 1994,
and there
is a distinct visual impression of a bar. This is clearly evident in the
near-infrared (see Fig. 7).

\begin{figure}
\resizebox{\hsize}{!}{\includegraphics{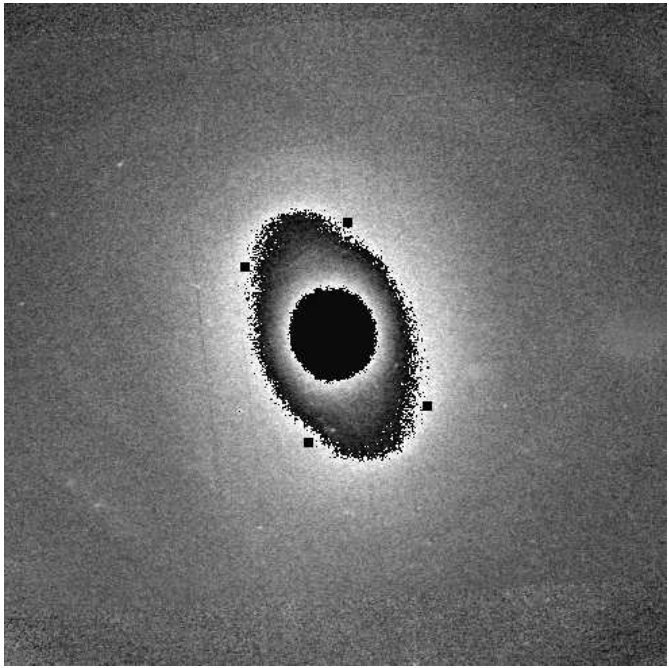}}
\caption{NGC 4450 is of de Vaucouleurs type SA. In the near-
infrared,
a bar of class 2 is identified, and our bar torque method finds the
locations (indicated by
four filled black squares) where the ratio of the tangential to the
mean
axisymmetric radial force reaches a maximum (in modulus), per
quadrant. The hint of a bar is also indicated optically, in Panel 110
of
`The Carnegie Atlas of Galaxies' by Sandage \& Bedke (1994).}
\label{make_mosaic_dot}
\end{figure}

 Similarly, Hubble category SB in the RSA has a wide range of
bar strengths. This category
commences at
a bar torque of class 1 (NGC 3344) and
reaches bar class 5 (e.g., NGC 7741) in Table 1 and bar class
6 in Table 1 of Buta \& Block (2001).
In other words, the bar strengths of some
RSA SB galaxies may be {\em weaker}
than those found in
RSA unbarred spirals such as NGC 1042 (Sc; near-infrared
bar class 3). This is not due to the uncertainties in the
$Q_b$ method, but instead reflects the difficulties of making
reliable bar strength judgments in the visual Hubble system. The
work of Knapen et al. (2000) reaches this
identical
conclusion, using their independent definition of bar strength.

\begin{table}
\caption{Mean bar torque versus visual bar classifications
for 64 RSA galaxies and 69 galaxies included in Appendix I
of de Vaucouleurs (1963). Galaxies in Table 1 of this paper
and in Table 1 of Buta \& Block (2001) which do not have bar
classifications from these sources are excluded from these
means. NGC 4618, a low-luminosity magellanic spiral, is also
excluded.}
\begin{tabular}{lcccc}
\hline\noalign{\smallskip}
Classification & $<Q_b>$ $\pm$ $\sigma$ &  $N$ & range \\
\hline\noalign{\smallskip}
         &            &    &           \\
RSA S    & 0.11$\pm$0.08 & 32 & 0.01--0.33\\
RSA SB   & 0.28$\pm$0.13 & 32 & 0.07--0.63\\
         &      &      &     &           \\
deV SA   & 0.06$\pm$0.04 & 14 & 0.01--0.14\\
deV SAB  & 0.16$\pm$0.08 & 32 & 0.02--0.33\\
deV SB   & 0.33$\pm$0.13 & 23 & 0.16--0.63\\
\hline\noalign{\smallskip}
\end{tabular}
\end{table}

The gravitational influence of bars is thus poorly recognized
by the Hubble classification scheme.
Table 2 shows that, on average, RSA SB galaxies have relative bar
torques
only 2.5 times as strong as in RSA S galaxies, with a very
large range in $Q_b$ in each class. In this table, two galaxies
of RSA type Sb/SBb have been included in the SB category.

(ii) Fig. 5 and Table 2 show that the situation is better for de
Vaucouleurs
classifications. The mean value of $Q_b$ changes smoothly with
de Vaucouleurs
family, and in fact verifies
the continuity in bar strength embodied in de
Vaucouleurs classifications. De Vaucouleurs SB galaxies have
relative
bar torques 5.5 times that of SA galaxies and twice that of SAB
galaxies. However, the scatter is still very large in the SAB and SB
categories.
In Table 1, SAB galaxies
encompass bar torque classes over the wide range 0 (e.g.,
NGC 6946) to 3 (e.g., NGC 4303), while SB galaxies encompass
the
range 2 (e.g., NGC 3351) to 5 (e.g., NGC 1300).
NGC 7479 (Buta \& Block \cite{buta01}) is a type SB galaxy of bar
class 6.

(iii) Fig. 6 shows that $Q_b$ correlates fairly well with
Martin's (1995) $(b/a)_{bar}$ parameter, confirming that bar
ellipticity
does provide a measure of bar strength. However, the scatter
at a given bar axis ratio is still large. Buta \&
Block (2001) had noted that highly elongated bars (such as in M83;
an example of Martin's bar ellipticity class 7) may have weak
torques.
At $(b/a)_{bar}$ = 0.5, $Q_b$ ranges from 0.1 to 0.5.
Thus, apparently strong bars with significant ellipiticity
(e.g., with elongations of $(b/a)_{bar}$  $\leq$ 0.5) may
be strong, weak, or intermediate as far as $Q_b$ torque values are
concerned. {\it $Q_b$ does not measure just the shape of an
isolated bar;
it also accounts for the disk in which the bar is embedded}.

(iv) We suspect that some of the scatter seen in Figs. 4 to 6 could
be
due to dilution of the bar torque by a strong bulge. We might expect
this because bars that are strong in terms of
the $m$=2 Fourier component of the optical
light distribution (Elmegreen \& Elmegreen \cite{elm85}) can have
either small or large relative torques,
depending on the relative mass of the bulge. If the bulge is
weak, then even a weak bar can have a strong torque
compared to the radial component of the force (e.g., NGC 1073).
Bars that are long can have a strong torque because the
end of the bar is far from the bulge (e.g., NGC 1300). This means
that
the simultaneous decreases in relative bulge strength and bar
length
with later Hubble type partially
offset each other, giving a relative torque that can either go up
or down, depending sensitively on the mass distribution.

However, when we replot Fig. 5 separating the points by Hubble
type, we find that
the galaxy-to-galaxy variation in the bar torque for each
bar type is not entirely the result of a varying force dilution from the
bulge. This is shown in Fig. 8, which includes the same
galaxies
as in Fig. 5 but with different symbols for early, intermediate, and
late Hubble types. These subtypes reflect a variation in the relative
strength of the bulge, with earlier types having stronger bulges in
both
barred and non-barred galaxies. Even within a subtype, some
optically
barred galaxies have smaller bar torques than some optically
unbarred
galaxies.
The relative bar torque comes from a mixture of bar amplitude,
radial profile and relative length, all combined with the bulge
strength.
These quantities vary in different ways along the Hubble subtype
sequence,
producing a wide range in relative bar torques.

\begin{figure}
\resizebox{\hsize}{!}{\includegraphics{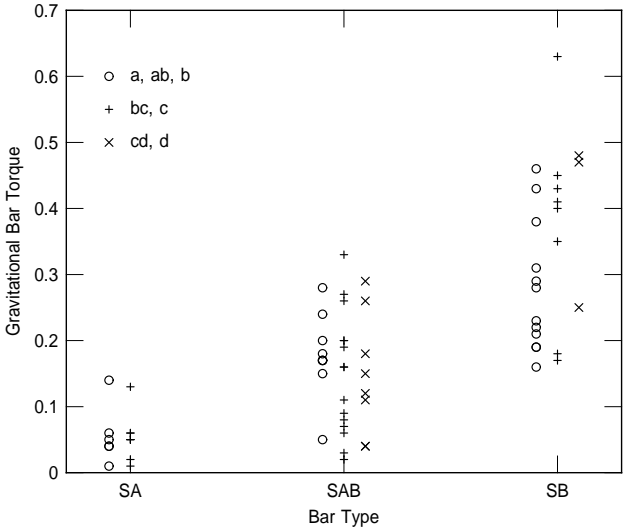}}
\caption{Relative bar torque versus
de Vaucouleurs bar type is shown with different symbols and a
slight
offset for the different Hubble subtypes. The wide range of
relative bar torques for each bar type is present for all
subtypes.}
\label{fig6.ps}
\end{figure}

Variations in torque with bar type are not obvious from the
morphology. If bars drive spirals, particularly in early Hubble types
where the presence of a bar correlates well with grand design spiral
structure (Elmegreen \& Elmegreen \cite{elm89}), they tend to do
so only
to the
point of saturation, producing very strong arms after only a few
revolutions. This is apparently true for both high and low bar
torques,
because even the low torques are enough to make strong spirals.
Thus
there is little sensitivity in spiral arm strength to the bar torque,
aside from the known sensitivity of arm strength to the relative
magnitude of the $m$=2 component of the infrared light (Elmegreen
\&
Elmegreen 1985).

Several bars in Fig. 2 show a two-component morphology: a
broad oval
bulge or bar-like structure extending out to about half or two-thirds
of the full bar length, and a thin spindle-like structure extending
out further. NGC 4314 is an example; the thick component is
outside
the ILR in this case because there is a small ILR ring much closer
to the center.  These two bar components generally appear to be
from
two distinct populations of stars: a warm or hot population to make
the thick bar, and a cool population to make the spindle. The two
components could also have formed at different times, with the hot
component being much older.
In this case, it would be interesting to study these galaxies as
possible examples where a relatively short bar formed first and
dissolved by the instabilities discussed in Hasan, Pfenniger, \&
Norman
(\cite{hasan93}), producing the oval we see today, and then another,
larger bar formed afterwards out of a younger population of stars and
gas,
producing the spindle.
We also note that some galaxies have
the
thick oval leading the thin spindle in the direction of rotation (e.g.,
NGC 1300), and other galaxies have it lagging (e.g., NGC 4123).
This
variation
might indicate some dynamical interaction between the two bars,
such as
an oscillation about the equilibrium aligned configuration.

\section{Robustness of $Q_b$ in the high-redshift universe}

Block et al. (2001) have suggested that a physically meaningful
classification system for high-redshift galaxies may be more easily
devised at rest-frame infrared wavelengths, rather than in the optical
regime. Sub-mm observations indicate that at least some of these
systems are heavily obscured by dust (Sanders 1999). Lessons
from our
local Universe are that
optical morphologies can be radically different from near-infrared
ones; some
optically flocculent
galaxies, for example, may have  beautiful {\it grand design} stellar
disks
when examined at $K$. The decoupling of gaseous and stellar
disks can
be dramatic (see e.g., Puerari et al. 2000, Elmegreen et al. 1999,
Grosb{\o}l \& Patsis 1998 and Block et al. 1994).
Block et
al. (2001) show that Fourier spectra may be generated on simulated
{\it Next Generation Space Telescope} (NGST) rest-frame $K$
post-stamp FITS images which may be as small as 1$''$ on a side.

\begin{figure}
\resizebox{\hsize}{!}{\includegraphics{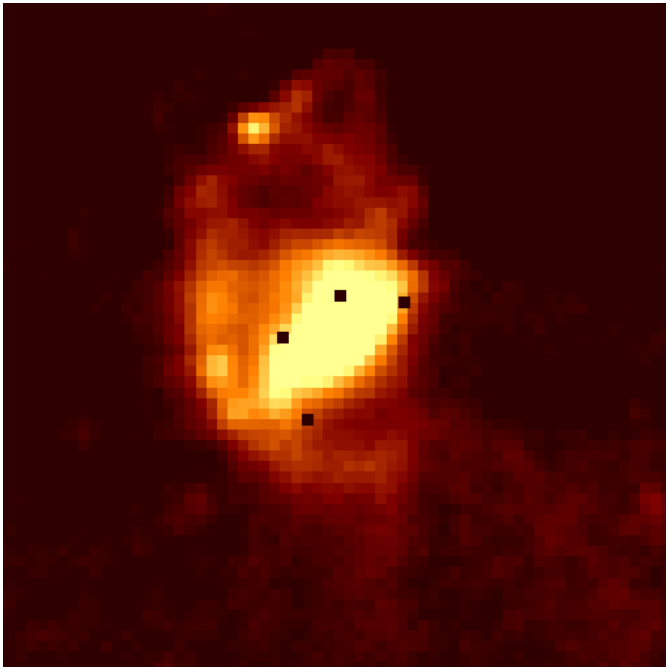}}
\caption{A simulated NGST image of the galaxy NGC~922 as it
would
  appear at a redshift of $z$ = 1.2, in its rest-frame $K$-band. The
  square post-stamp
 FITS image measures $\sim$ 2$''$ on a side.
The image
is recreated from a groundbased
  $K$-band image obtained at the NASA Infrared Telescope
Facility, following the simulation methodology of Takamiya
  (1999).
At a redshift of $z$ = 1.2,
  the  characteristic `bar signature' (indicated by the pattern of
  `four dots' as in Fig. 2) is
 still found in a robust manner,
  even though by eye the bar region has suffered considerable
  degradation as a result of limited spatial resolution. In recreating
  this
  simulated NGST $M$-band (4.7$\mu m$) image, an exposure
time of
  one-hour is assumed. For further details, see Block et al. (2001),
  where a groundbased $K$-band image of NGC~922 may also be
found.}
\label{fig7.ps}
\end{figure}

With the tremendous importance in attempting to bridge the low and
high-redshift universe from a morphological point of view, it is
natural to enquire how robust our bar torque method
is as we degrade the quality
of the images. First indications are that $Q_b$ is surprisingly robust
 in the presence of
noise and limited spatial resolution.

Recreated rest-frame
$K$-band NGST images of the galaxies
NGC~2997 and NGC~922 when moved out to
redshifts of $z$ = 0.7 and $z$ = 1.2, are presented in Block et
al. (2001).
We have applied our $Q_b$ method to NGC~922,
which may serve as an excellent morphological interface between
the
low and high redshift universe (see Block
et al. 2001 for full details).

The results are indicated in Fig. 9. Even in the
presence of significant image degradation of NGC~922  when this
galaxy
is moved out to
a redshift of $z$=1.2 (the eye now cannot
easily distinguish the boundaries of a bar, see Fig. 9) --
the four locations where the ratio of the
tangential force to the mean axisymmetric radial force reaches a
maximum (in modulus) may readily be identified. NGC~922 in the
groundbased as well as in the simulated image (Fig. 9) belongs
to
bar class 2. Further work on moving each galaxy in Table 1 out
to
$z$ $\sim$ 1 ($L$ and
$M$ band imaging) as well as to $z$ $\sim$ 3
(simulated $N$-band imaging) at rest-frame $K$-band, with NGST, are in
progress. This would
yield a statistically complete sample upon which to test the
robustness in preservation of bar torque class with increasing $z$.

\section{Conclusions}
\label{conclusion}

We have shown that RSA and de Vaucouleurs bar classifications
correlate
poorly with relative bar torque. It is not unexpected
that there would be some {\it average} correlation; as shown in
Table 2,
the S, SA/SAB categories do indeed select weaker bars on
average than
the
SB categories. However, the scatter in any given category is so
large
that
all categories have significant overlap. Some RSA SB galaxies have
much weaker bars than some RSA S galaxies. Deprojected bar
axis
ratios
are a better measure of bar
strength, but even with this more quantitative parameter, the scatter
in $Q_b$ is still large for each axis ratio.

The relative bar torque parameter $Q_b$ provides a logical addition
to
the dust-penetrated classification of Block \& Puerari (1999). As
noted
by Buta \& Block (2001), there is still room for improvement that
would
involve better estimates of orientation parameters, more quantitative
treatments of the bulge, better tests of the constant M/L
assumption
using near-IR colors, and a derivation of the vertical scaleheight as
a fraction of the radial scalelength. These issues will be addressed
in
more depth in a future paper.

At least for NGC~922, the $Q_b$ bar torque method is surprisingly
robust
 in the presence of
noise and limited spatial resolution, when simulated images of this
 galaxy
are recreated at a redshift of $z$ = 1.2 in the dust-penetrated
 rest-frame
$K$-band.

\begin{acknowledgements}

We thank the anonymous referee for valuable comments.
A note of deep gratitude is expressed by DLB to the entire SASOL
Board, including
P. Kruger (Chairman) and  P. Cox (CEO).
SASOL Corporate Affairs Manager Mr S. Motau has been most
helpful.
We are also most indebted to the
Anglo American Chairman's Fund; in particular, we warmly thank
the
Board of Trustees and CEO Mrs. M. Keeton for the funding of two
of us
(RJB and IP) to work in South Africa. DLB freely acknowledges the
saving grace of God in allowing this research to continue after a
motor car accident; `Iam ediximus deum universitatem hanc mundi
verbo
et ratione et virtute molitum' (Tertullian).
We thank R.S. de Jong and D.M.
Bramich for
help with
the new images. IP would like to express his deep gratitude to 
E. Athanassoula for interesting discussions on spiral structure. These
discussions were made possible in terms of the ECOS/ANUIES exchange project
M99-U02, for which IP is most grateful. The groundbased near-infrared
image 
of NGC~922
from
which Fig. 9 was created, was kindly secured by Dr A. Stockton.

\end{acknowledgements}

{}


\begin{thebibliography}{}


\bibitem[2000]{abr00}
    Abraham, R. \& Merrifield, M.R. 2000, AJ 120, 2835


\bibitem[1992]{ath92}
   Athanassoula, E. 1992, MNRAS 259, 328


\bibitem[1991]{block91}
   Block, D.L. \& Wainscoat, R.J. 1991, Nat 353, 48


\bibitem[1994]{block94}
   Block, D.L. et al. 1994, AA 288, 365`

\bibitem[1999]{block99}
    Block, D.L. \& Puerari, I. 1999, AA 342, 627

\bibitem[1999]{mask00}
    Block, D.L. et al. 1999,
    Ap\&SS, 269, 5

\bibitem[2001]{blo01}
    Block, D.L. et al. 2001, A\&A (in press)


\bibitem[2001]{buta01}
    Buta, R.J. \& Block, D.L. 2001,
    ApJ, 550, 243

\bibitem[1996]{buta96}
Buta, R.J. \& Combes, F. 1996, Fund. Cosmic Phys. 17, 95

\bibitem[1981]{combes81}
   Combes, F. \& Sanders, R.H. 1981, A\&A, 96, 164

\bibitem[1918]{curtis18}
 Curtis, H. D. 1918, Pub. Lick Obs. XIII, Part I, 11

\bibitem[1998]{deb98}
    Debattista, V. \& Sellwood, J.A. 1998, ApJ, 493, 5

\bibitem[2000]{deb00}
    Debattista, V. \& Sellwood, J.A. 2000, ApJ, 543, 704


\bibitem[1998]{deg98}
     de Grijs, R. 1998, MNRAS, 299, 595


\bibitem[1959]{dev59}
    de Vaucouleurs, G. 1959, Handbuch der Physik, 53, 275


\bibitem[1963]{dev63}
    de Vaucouleurs, G. 1963,
    ApJS 8, 31

\bibitem[1985]{elm85}
    Elmegreen, B.G. \& Elmegreen, D.M. 1985,
    MNRAS 288, 438

\bibitem[1987]{elm87}
    Elmegreen, D.M. \& Elmegreen, B.G. 1987, ApJ 314, 3


\bibitem[1989]{elm89}
    Elmegreen, B.G. \& Elmegreen, D.M. 1989,
         ApJ 342, 677


\bibitem[1999]{elm99}
    Elmegreen, D.M. et al. 1999, AJ 118, 2618

\bibitem[2000]{esk00}
    Eskridge, P. et al. 2000, AJ 119, 536

\bibitem[1992]{fre92}
      Freeman, K.C. 1992, in Physics of Nearby Galaxies: Nature or
Nurture?
      eds. T.X. Thuan, C. Balkowski \& J. Tran Thanh, Gif-sur-Yvette,
      Editions Frontiere, 201


\bibitem[1983]{gil83}
   Gilmore, G. \& Reid, N. 1983, MNRAS 202, 1025


\bibitem[1998]{gros98}
    Grosb{\o}l, P.J. \& Patsis, P.A. 1998, AA 336, 840


\bibitem[1993]{hasan93}
     Hasan, H., Pfenniger, D. \& Norman, C. 1993, ApJ 409, 91

\bibitem[1926]{hubble26}
    Hubble, E. 1926,
   ApJ 64, 321


\bibitem[1981]{kennicutt81}
   Kennicutt, R.C. 1981, AJ, 86, 1847


\bibitem[1999]{kna99}
Knapen, J.H. 1999, in The Evolution of Galaxies on Cosmological
Timescales, J.E. Beckman \& T.J. Mahoney, Eds., ASP Conf. Ser.
187, 72


\bibitem[2000]{knapen20}
    Knapen, J.H., Shlosman, I. \& Peletier, R.F. 2000, ApJ, 529, 93


\bibitem[1979]{kormendy79}
     Kormendy, J. 1979, ApJ, 227, 714


\bibitem[1995]{mar95}
   Martin, P. 1995, AJ 109, 2428






\bibitem[1996]{per96}
  Persic, M., Salucci, P. \& Stel, F. 1996, MNRAS, 281, 27


\bibitem[2000]{puerari2000}
   Puerari, I. et al. 2000, AA 359, 932


\bibitem[1994]{qui94}
Quillen, A. C., Frogel, J. A., \& Gonz\'alez, R. 1994, ApJ, 437, 162


\bibitem[1997]{reg97}
   Regan, M.W. \& Elmegreen, D.M. 1997, AJ 114, 965


\bibitem[1998]{rozas98}
    Rozas, M., Knapen, J.H. \& Beckman, J.E. 1998, MNRAS 301,
631

\bibitem[1961]{san61}
Sandage, A. 1961, ``The Hubble Atlas of Galaxies'' (Carnegie Inst.
of Wash. Publ. No. 618)

\bibitem[1981]{san81}
    Sandage, A. \& Tammann, G. 1981, ``A Revised Shapley-Ames
Catalog
    of Bright Galaxies'' (Carengie Inst; Wash. DC)

\bibitem[1994]{sandage94}
    Sandage, A. \& Bedke, J. 1994, ``The Carnegie Atlas of
Galaxies''
    (Carnegie Inst; Wash. DC)



\bibitem[1999]{mask00}
    Sanders, D.B. 1999,
    Ap\&SS, 269, 381



\bibitem[1994]{schroder94}
    Schr\"oder, M.F.S. et al. 1994, A\&AS, 108, 41


\bibitem[1993]{sel93}
       Sellwood, J.A. \& Wilkinson, A. 1993, Rep. Prog. Phys. 56,
173


\bibitem[2001]{sted01}
    Stedman, S. \& Knapen, J.H. 2001, Ap\&SS, in press.


\bibitem[1999]{tak99}
    Takamiya, M. 1999, ApJS, 122, 109

\end{thebibliography}
\end{document}